# Gearshift Fellowship: A Next-Generation Neurocomputational Game Platform to Model and Train Human-AI Adaptability
## *Toward an Ecosystem for Neuroscientific Discovery, Clinical Phenotyping, and Personalized Meta-Learning*

Nadja R. Ging-Jehli[1], Russell K. Childers[2], Joshua Lu[3], Robert Gemma[3], Rachel Zhu[4]

[1] Carney Institute for Brain Science, Brown University, Providence RI 02906, USA
[2] BGBehavior LLC, Columbus OH 43214, USA
[3] Center for Computation and Visualization, Brown University, Providence RI 02906, USA
[4] Rhode Island School of Design, Providence RI 02906, USA

Corresponding Author: Nadja R. Ging-Jehli, PhD; Carney Institute for Brain Science & Department of Cognitive and Psychological Sciences, Brown University, Providence, RI (USA). nadja@gingjehli.com.

**Abstract.** How do we learn when to persist, when to let go, and when to shift gears? *Gearshift Fellowship* (GF) is the prototype of a new *Supertask* paradigm designed to model how humans and artificial agents adapt to shifting environment demands. Grounded in cognitive neuroscience, computational psychiatry, economics, and artificial intelligence, *Supertasks* combine computational neurocognitive modeling with serious gaming. This creates a dynamic, multi-mission environment engineered to assess mechanisms of adaptive behavior across cognitive and social contexts. Computational parameters explain behavior and probe mechanisms by controlling the game environment. Unlike traditional tasks, GF enables neurocognitive modeling of individual differences across perceptual decisions, learning, and meta-cognitive levels. This positions GF as a flexible testbed for understanding how cognitive-affective control processes, learning styles, strategy use, and motivational shifts adapt across contexts and over time. It serves as an experimental platform for scientists, a phenotype-to-mechanism intervention for clinicians, and a training tool for players aiming to strengthen self-regulated learning, mood, and stress resilience. Online study (n = 60, ongoing) results show that GF recovers effects from traditional neuropsychological tasks (construct validity), uncovers novel patterns in how learning differs across contexts and how clinical features map onto distinct adaptations. These findings pave the way for developing in-game interventions that foster self-efficacy and agency to cope with real-world stress and uncertainty. GF builds a new adaptive ecosystem designed to accelerate science, transform clinical care, and foster individual growth. It offers a mirror and training ground where humans and machines co-develop together deeper flexibility and awareness.

# 1 Introduction

*How do we learn to shift gears – when to persist, when to let go, and how to adapt in ways that preserve our openness to explore new opportunities while staying grounded in purpose and agency?* In a time marked by rapid change and rising uncertainty – economically, socially, and technologically – this question has never been more pressing. It cuts across disciplines from neuroscience and behavioral economics to mental health, artificial intelligence (AI), and education. Adaptive behavior is the capacity to flexibly adjust thoughts, actions, and strategies in response to changing environmental demands [22]. It is central to the study of brain function [14, 52], human mental health [30, 54], and the development of intelligent systems [4, 46]. In neuroscience, adaptability is linked to the coordination of distributed neural systems involved in inference, control, and motivation. Disruptions in these mechanisms are observed across a range of psychiatric conditions, where individuals often become stuck into rigid, maladaptive patterns of thought and behavior [2, 30, 54, 60]. At the same time, artificial intelligence systems, even those built on powerful learning algorithms, frequently struggle to generalize beyond narrowly trained contexts; particularly when task structure and purpose shifts [12, 46]. These converging limitations underscore the need for unified frameworks that can elicit, model, and compare adaptive processes across both biological and artificial agents, at behavioral, motivational, and neurocomputational levels of analysis.

Despite growing interest in adaptability across disciplines, we still lack environments that can systematically probe the mechanisms behind flexible thinking and behavior while remaining scalable, ecologically valid, and computationally tractable (i.e., capable of explaining behavior, rather than merely predicting future outcomes). This gap is especially pressing as mental health needs rise, AI systems grow more complex, and science faces increasing pressure to deliver frameworks that can translate across labs, clinics, and real-world settings. Serious games offer a promising avenue to bridge this divide, but most implementations do not capture the dynamic, multi-layered nature of adaptation [1, 32, 51, 53, 57]. Existing implementations have successfully improved engagement and accessibility, particularly in mental health, education, and training contexts. Yet, few are explicitly designed to probe the underlying mechanisms of adaptability [32, 51]. As a result, their potential to support scientific discovery, clinical insight, or human-AI co-evolution remains underexplored [15]. We argue that when paired with formal economic decision theory and computational modeling from computational psychiatry, neuroscience, and AI, serious games can evolve into powerful platforms for dissecting adaptive brain mechanisms to optimize them for enhancing productivity, identifying digital markers of flexibility, predicting behavioral and clinical outcomes, and guiding support that is both just-in-time and adaptive.

We introduce *Gearshift Fellowship* (*GF*), a computationally engineered serious game platform designed to assess, model, and train adaptive behavior across shifting task environments. The platform is grounded in theories of meta-learning, controllability, and decision-making, and supports deployment in both research and applied settings. By integrating hierarchical computational analyses, serious gaming, and AI-informed personalization, GF aims to establish a new ecosystem that redefines how scientists, healthcare providers, industry partners, and participants collaborate and co-evolve. It offers a unified, user-centered framework for discovering neurocomputational mechanisms, digitally phenotyping mental rigidity and maladaptive behavior that can complement clinical care, and deliver personalized, adaptive training to strengthen self-efficacy and agency in coping with real-world stress, uncertainty, and mood fluctuations. As such, we position GF at the intersection of translational serious gaming, computational cognitive-affective modeling, and neuroscientific decision theory. In what follows, we first situate our work in relation to prior approaches, then introduce a novel conceptual paradigm referred to as *Supertask*, which reframes serious games as evolving environments for probing and training adaptability (when being integrated with neuroeconomic theory). We then present GF's architecture, describe its implementation, and offer first application insights before discussing future directions.

## 2 Related Work

Serious games are increasingly used across mental health, education, and cognitive neuroscience to assess behavior in ecologically valid yet structured ways [1, 49, 53]. Traditional serious games have primarily focused on engagement and user acceptability, with limited emphasis on formal modeling of adaptive behavior [23, 43]. Meanwhile, neuropsychological (gamified) tasks used in computational psychiatry often prioritize experimental control over ecological validity and fail to provide direct value to the user [1, 9, 30]. Efforts to bridge these domains, such as gamified reinforcement learning tasks [66] or adaptive assessment tools [64, 68, 69], typically remain static, domain-specific, or limited in scope. In AI, benchmarks for meta-learning and task inference have gained traction [34, 44], but few environments allow direct comparison or co-evolution between human and artificial agents under shared conditions.

Recent platforms such as OpenMind [45], AI Gym environments [8], and digital phenotyping apps [55] each address aspects of learning or adaptation, but lack an integrated, multi-user architecture that supports scientific modeling, clinical assessment, and individualized gameplay.

GF advances this landscape as it is based on a new Supertask paradigm – a platform where adaptive behavior can be assessed, trained, and co-evolved across humans and intelligent systems. Beyond performance metrics, GF fosters co-evolution between human players and adaptive agents, cultivating new levels of self-awareness, strategic flexibility, and conscious control. By integrating neurocomputational modeling with insights from behavioral economics and AI, GF enables a deeper scientific understanding of mental and behavioral adaptation. This, in turn, sets the stage for tools that can enhance not only real-world mental resilience (coping with stress and fluctuations in mood, attention, and mental clarity) but also enhance adaptive learning, productivity and conscious decision-making under uncertainty. In this way, GF serves not just as a platform, but as the foundation for a new adaptive ecosystem – one where science, care, and human potential evolve together.

## 3 Theoretical Foundation

Understanding and enhancing human adaptability, our capacity to adjust thoughts, actions, and goals in response to uncertainty or shifting demands, is a central concern across neuroscience, psychiatry, education, and artificial intelligence [1, 6, 15, 26, 30, 52]. However, adaptability is not a monolithic construct: it unfolds through the interaction of multiple computational and motivational mechanisms operating at different timescales, such as inference of latent task structure, adjustment of learning rates, and shifts in control allocation to guide attention during decision-making [26, 28, 62]. GF is built to isolate and probe these components through gameplay that elicits structured variation in context, reward framing, effort demands, and risk.

At its core, the platform is informed by computational models of sequential sampling [25, 58], hierarchical reinforcement learning [6, 21], Bayesian inference [33], and meta-learning [24, 65]. These models view agents (human or artificial) as dynamically optimizing not only for external reward, but also for control, uncertainty reduction, and meaningful structure discovery. For example, when faced with changing goals or ambiguous feedback, successful adaptation often involves meta-cognitive processes such as updating beliefs about task rules or switching strategies [48, 61]. These capacities are disrupted in many mental health conditions [13, 39], yet remain difficult to to measure in traditional neuropsychological tasks, and hard to cultivate in real-world settings [27, 56].

Unlike standard games or fixed cognitive batteries, GF is explicitly designed to elicit *meta-learning behavior*: the ability to learn how to adjust decision, reasoning, and learning strategies across contexts with shifting cognitive and social demands that vary in controllability of different aspects of uncertainty. The platform leverages variable mission structures that systematically shift the relevance of past knowledge, thus requiring players to infer when to persist vs. when to explore. This design enables the quantification of individual differences in task inference, effort-reward trade-offs, and controllability-seeking behaviors – concepts that are clinically and neuroscientifically significant but have traditionally been studied in isolation. GF relates these concepts to each other, deconstructing them within one computational framework to relate them to each other, aligning with emerging theories in AI and cognitive neuroscience.

In contrast to single-purpose tasks or fragmented mini-games, GF introduces a novel testbed we call a *Supertask*. That is, a unified paradigm engineered to assess and train adaptability across shifting demands, timescales, and contexts. By embedding these theoretical constructs into intuitive game mechanics, GF offers for scientists a testbed for developing and validating mechanistic models of adaptability in both typical and clinical populations, and for comparing them to AI agents exposed to the same structured environments.

## 4 The Supertask Paradigm

In clinical neuroscience and computational psychiatry, many paradigms have been developed to isolate mechanisms underlying behavioral rigidity such as avoidance behavior, effort–reward sensitivity, or intolerance to uncertainty [29, 30, 50, 54, 60]. While these paradigms have advanced our understanding of specific processes, they are typically studied in isolation, lack ecological validity, and are not designed to capture how these mechanisms interact dynamically across contexts [1, 30, 38, 59]. Hence, we still have limited tools to assess and model how people adapt (or fail to adapt) across shifting conditions that characterize real-world goal pursuit; and how this relates to symptoms of depression, anxiety, and attention-deficit/hyperactivity disorder (ADHD).

To address this, we introduce a new conceptual and structural paradigm to which we refer to as the Supertask: a mission-based game environment designed to study and model a range of cognitive, affective, and motivational processes (e.g., learning, decision-making, and behavioral regulation) across structured, reconfigurable contexts. A Supertask consists of a sequence of distinct yet structurally related missions that vary in framing, goal structure, reward contingencies, and cognitive demands. Because these missions are embedded within a unified architecture, Supertasks support both systematic manipulation and cross-context modeling. This design enables the study of task-specific behavior and overarching patterns of flexibility, while also relating them to distinct neuropsychological phenomena such as avoidance behavior, reward sensitivity, and effort-based decision-making. Crucially, Supertasks are designed to support unified hierarchical computational modeling of the generative dynamics of behavior across missions and timescales, enabling the joint analysis of both state-dependent and trait-level characteristics. This is possible because gameplay is hierarchically organized: at the nano-level, players engage in different decisions within trials that are nested in blocks at the micro-level, which in turn are nested into missions and longitudinal assessments that allow for predictions of real-world outcomes (stress, mood). Importantly, the underlying sequence of actions towards a goal (i.e., the core game mechanic) is consistent across trials, blocks, and missions. Though, missions differ in framing and contingency structure, requiring meta-learning and strategic updating.

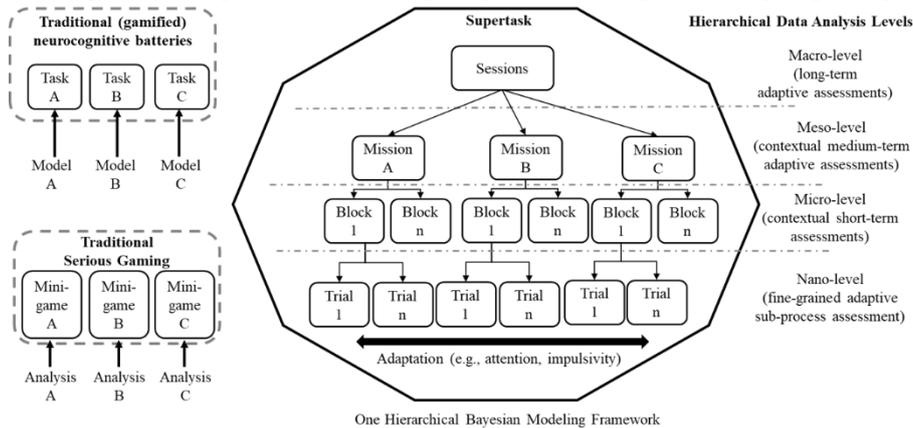

**Fig. 1.** Conceptual overview of the Supertask paradigm. Traditional paradigms often consist of loosely connected tasks or games assembled into batteries that are separately analyzed. This fragmented approach frequently yields weak correlations across domains, making it difficult to determine whether these dissociations reflect true domain

specificity or arise from uncontrolled differences in task structure, demands, and context. Instead, the Supertask paradigm organizes behavior within a unified, hierarchical framework with the same underlying core game mechanism. The Supertask consists of nested missions that span meso-, micro-, and nano-level challenges, enabling assessment of behavior across multiple timescales and contextual layers. Each level engages distinct cognitive and motivational processes, facilitating integrated analysis of learning, decision-making, and generalization. This layered organization supports adaptive learning, context inference, and performance tracking across multiple timescales and contexts. Artificial agents and human players can be compared and co-evolved within one structure, supporting computational benchmarking and translational relevance.

While Supertasks draw inspiration from behavioral economics, cognitive control and reinforcement learning paradigms (e.g., reversal learning [17], effort discounting [5], or the two-step task [20]), they go beyond traditional experimental designs in both scope and structure. Rather than isolating a single construct, Supertasks allow for the integration and dissociation of multiple cognitive and motivational phenomena within a single, evolving task environment. Compared to gamified cognitive batteries like Lumosity [36] or Cambridge Brain Sciences [35], Supertasks are designed not for screening but for computational inference, and emphasize ecological validity without sacrificing experimental control. In the AI domain, Supertasks share common ground with meta-learning and multi-task training environments such as BabyAI [11], Meta-World [67], and OpenAI Gym [8], but they are uniquely designed to support shared task structures between humans and artificial agents, enabling direct comparisons in adaptive behavior. Finally, while serious games like Foldit [18] and Sea Hero Quest [19] have demonstrated the potential of gameplay for large-scale data collection and cognitive mapping, they do not provide the hierarchical task structure, model-driven adaptability, or integrative design that define the Supertask paradigm. In this way, Supertasks represent a new class of experimental environments that are computationally grounded, behaviorally rich, and designed to probe the interaction of multiple cognitive and motivational processes over time and context. The goal of such an environment is to support co-evolution of humans and artificial agents.

By embedding this structure within a serious game platform, Supertasks offer also a powerful new framework for translational research. They maintain experimental control while achieving greater ecological validity, enabling both the modeling of adaptive processes and the delivery of personalized feedback. Ultimately, Supertasks are not just tools for assessment. Instead, they are flexible, generative environments that can support learning, clinical insight, and human-AI co-evolution within a single adaptive system. GF operationalizes the Supertask paradigm through a modular game architecture composed of contextually varied missions embedded within a continuous narrative arc. The platform is built to balance three competing demands: 1. ecological validity and engagement; 2. experimental control and interpretability; and 3. computational tractability for mechanistic modeling and integrating adaptive artificial agents. As such, GF represents not only an instance of the Supertask paradigm, but a new class of adaptive game environments for supporting human-computer co-evolution.

> A Supertask is a mission-based game environment designed to study and model a range of cognitive and motivational processes (e.g., learning, decision-making, and behavioral regulation) across structured, reconfigurable contexts. It relates distinct neuropsychological phenomena (e.g., avoidance behavior, reward sensitivity, effort-based decision-making) within a hierarchical computational framework that captures the generative dynamics of behavior across missions and time. It supports the modeling of state- and trait-level characteristics and has five properties:
> 1. **Goal-Directed Behavior**: Players pursue evolving goals through a sequence of interdependent decisions. Each mission requires the execution of multiple actions in service of higher-order objectives.
> 2. **Contextual Reconfiguration**: While the core game mechanic remains consistent, each mission shifts contextual variables such as framing, goal structure, reward contingencies, and cognitive demands. This enables systematic manipulation of adaptation-relevant parameters within a unified environment.
> 3. **Hierarchical Embedding**: Trial-wise decisions contribute to block-level and mission-level outcomes, allowing for modeling of patterns at multiple levels. This promotes the study of abstraction, generalization, and transfer.
> 4. **Continuity and Memory Dependence**: Player behavior is influenced by prior choices and accumulated performance history. Missions are interdependent, requiring dynamic trade-offs between different strategies.
> 5. **Unified Computational Modeling**: The consistency of core mechanics across missions supports formal modeling within a single computational framework, enabling joint inference of variables such as control, uncertainty, effort sensitivity, and motivational dynamics over time.

**Box 1.** Definition and core properties of a Supertask paradigm. A summary of the defining features of the Supertask paradigm, which integrates structured variability, goal-directed behavior, and unified computational modeling to support the study of cognitive and motivational processes across dynamic task environments.

## 5 Platform Architecture of Gearshift Fellowship (GF)

GF is the flagship of the first implementation of a Supertask paradigm focused on adaptive behavior under varying levels of uncertainty and controllability across cognitive and social learning missions. It is a multi-mission serious game that combines car-based navigation, detective-style reasoning, and decision-making under uncertainty within an integrated hero's journey to assess and train behavioral adaptability. Each mission is designed to probe distinct decision-making signatures, while the overall arc of gameplay captures higher-order dynamics, including meta-learning, strategy flexibility, and motivational shifts over time. Players pursue evolving goals through context-sensitive decisions, requiring them to adapt not only within each mission but also to infer higher-order patterns (e.g., controllability of outcomes) and strategically adjust their behavior across missions. This multi-level structure enables the disentangling of shared and distinct mechanisms underlying adaptive versus rigid behavior.

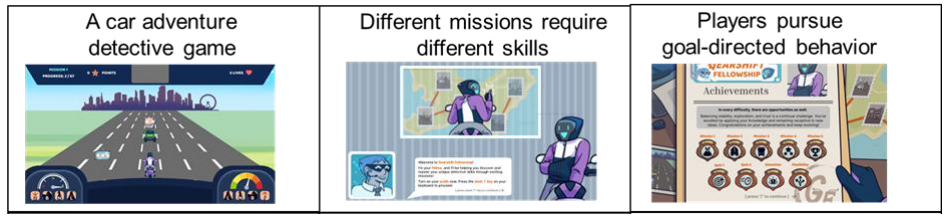

**Fig. 2.** Game Design and Features of GF. Players assume the role of motorcycle detectives chasing target vehicles. To succeed, they must evaluate encoded information, identify rewarding opportunities, and avoid threats across missions that vary in effort, uncertainty, and controllability. Each mission is grounded in a shared core mechanic, with underlying structures designed for cognitive modeling of adaptive behavior. Over time, the platform supports dynamic reconfiguration based on gameplay patterns to create personalized profiles of adaptive learning.

### 5.1 Modular & Hierarchical Architecture

GF is built upon a shared, hierarchical scaffolding structured at the trial, block, and mission levels to probe distinct cognitive and motivational mechanisms relevant to clinical, neurocognitive, and behavioral economic theories of adaptability. Each mission targets a specific psychological domain (e.g., cognitive flexibility, social learning) that is traditionally studied in isolation (Fig. 3, left). Within missions, blocks serve as intermediate contexts that systematically vary latent task structures (e.g., uncertainty, reward contingencies), probing players' adaptability along these dimensions. Although each mission is self-contained and intuitive to the player, all share a common core mechanic and contribute to an integrated gameplay history. This modular yet consistent structure enables both targeted manipulation of task features and unified, cross-context modeling of adaptive behavior.

As shown in Fig. 3 (right), players assume the role of motorcycle detectives tasked with chasing target vehicles and decoding their codes (shown in the gray pedestal) to collect reward packages and avoid traps. Fig. 3 shows that the core mechanic remains consistent across missions, while the underlying cognitive demands shift to probe different aspects of adaptability.

In Mission 1, players are explicitly told the higher-order rule associated with each car (e.g., whether to attend to a letter or a number), allowing them to focus on efficiently encoding and applying the relevant feature. This mission functions as a cued task-switching setting. In Mission 2, the higher-order rule must be learned through experience, creating a hierarchical reinforcement learning setting. In Mission 3, a social dimension is introduced. Players must decide whether to independently solve the code or allow a bystander (another "driver") to intervene and solve it on their behalf. These partners follow different (sometimes suboptimal) strategies, requiring players to reason and learn whom to trust, whom to ignore, and when to assert control, probing social inference, risk evaluation, and adaptive trust at meta-cognitive levels. Hence, in addition to individual cognitive processes, GF integrates interactive paradigms rooted in game theory to examine social decision-making. These interactions provide a framework to study belief formation, learning, and strategy use in social contexts. This links cognitive, social-cognitive, and affective processes. While clinically significant, these dimensions remain underexplored in scalable digital environments [30, 59]. GF offers novel opportunities to capture them within a unified computational framework.

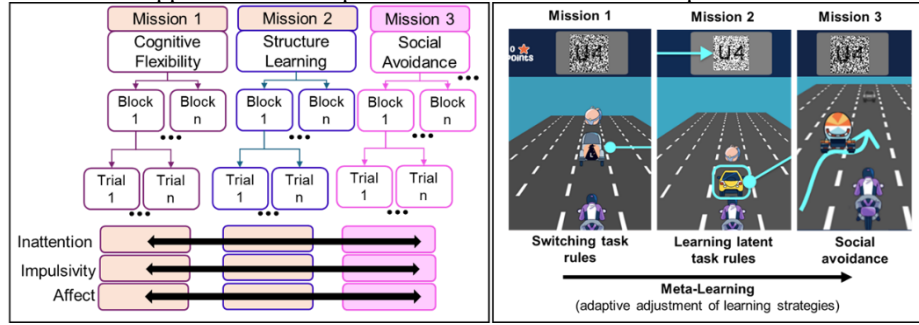

**Fig. 3.** A snapshot of the hierarchical game structure of GF. *Left.* Each mission targets a distinct clinical or psychological construct (e.g., avoidance behavior) that is typically studied in isolation. Trials are nested within blocks, which are nested within missions, dissecting the mechanisms that give rise to these constructs and examine their malleability across timescales: short-term (trials), medium-term (blocks), and longer-term (across missions). *Right.* All missions follow a structurally consistent trial–block–mission format, despite targeting different constructs. This structural alignment enables unified hierarchical modeling and supports cross-mission comparisons, facilitating the inference of individual-level phenotypes of adaptability.

### 5.2 Computational Modeling Framework

GF's computational back end enables model-based analysis of players' gaming patterns. The game continuously tracks trial-level decisions, response times, and internal state transitions, providing rich data streams for formal cognitive modeling. These data are analyzed using unified computational models (e.g., reinforcement learning, Bayesian inference, and sequential sampling models) that approximate the generative processes underlying the game behavior (Fig. 3, left). Because missions share consistent structural scaffolding, these models can be applied across tasks, allowing researchers to dissociate generalizable cognitive strategies from task-specific responses. Critically, model parameters are interpretable in psychological terms, capturing processes such as inference precision, learning asymmetries, perceptual biases, or strategic updating. Past research has shown that these models uncover latent mental processes not directly observable from behavior alone and make it possible to quantify individual differences in adaptability and cognitive rigidity across multiple levels, from perceptual inference to meta-cognitive control. Over time, this approach supports the generation of individualized cognitive profiles and the longitudinal tracking of behavioral change. It can also be integrated with physiological measures (e.g., electroencephalography, eye-tracking), enabling novel neuroscientific discoveries and the identification of bio-computational markers that can further inform personalized medicine.

### 5.3 Target Audience Design Considerations

To ensure accessibility and engagement across diverse populations, including reluctant players, clinical users, and non-gamers. GF was designed with a focus on intuitive interaction, motivational scaffolding, and adaptive feedback. The story is culturally accessible and lighthearted, framing the hero's journey in a non-judgmental, relatable way. This stands in contrast to traditional neuropsychological tasks, which can be abstract and complex for clinical populations, making GF more approachable and engaging. For individuals with low intrinsic motivation, gameplay is structured around meaningful missions with clear goals and progressive challenge pacing, while minimizing cognitive overload (Fig. 2). Clinical populations, including those with attentional or emotional dysregulation, are supported through simplified visuals, short mission durations, and instant feedback that reinforces agency and success. The interface is deliberately minimalistic and non-competitive, using clear affordances and calm framing to reduce anxiety. Tutorials and in-game prompts are delivered gradually, fostering a sense of safety and control while still enabling the collection of rich behavioral data.

## 6 Behavioral Results from the First Three Missions

To illustrate GF's capacity to capture individual differences in adaptive behavior, we conducted an online study with a normative adult sample (N = 60; ages 18–40; 50% female, 50% male; data collection ongoing). Participants completed a one-hour gameplay session consisting of five missions, each designed to probe distinct cognitive mechanisms such as rule switching (Mission 1), instrumental learning (Mission 2), and risk–reward trade-offs under social uncertainty (Mission 3). Participants completed also standardized self-report questionnaires administered via Qualtrics, including measures of self-efficacy [10] and symptoms of depression, anxiety, and stress, assessed via the Depression Anxiety Stress Scales (DASS-21) [37]. Participants also completed self-report measures of attention and impulsivity (e.g., ADHD symptom scales [16]), which are beyond this study's scope but will inform future clinical applications.

We next present descriptive behavioral results from the first three missions to demonstrate the platform's construct validity (i.e., replicating findings from classical neuropsychological tests) as well as its sensitivity to individual variation in cognitive flexibility, learning, and adaptive social inference. The analyses below provide an initial demonstration of GF's ability to elicit structured behavioral variation across distinct cognitive domains. Rather than exhaustively analyzing the full dataset or model space, we present descriptive behavioral results from the first three missions to illustrate how GF as a first prototype of a Supertask paradigm captures both canonical effects and meaningful individual differences. Full computational modeling and additional mission analyses will be reported in future work.

### 6.1 Mission 1: Task-Switching and Cognitive Flexibility

Mission 1 served as a cued rule-application task, tapping into cognitive flexibility through task-switching. This means that participants were given the higher-order rule (i.e., which feature of the code to attend to) via car-based cues. Participants showed classic switch costs (Fig. 4A): response times (RTs) were significantly slower on switch compared to no-switch trials ($\Delta$RTs = 525 ms, SE = 66 ms, $p < 0.001$), and accuracy was significantly lower on switch trials ($\Delta$Accuracy = -0.14, SE = 0.02, $p < 0.001$). These effects replicate well-established findings from cognitive task-switching paradigms [30, 31, 42]. Importantly, individual differences were associated with these performance patterns. Specifically, higher self-reported stress correlated with greater switch costs ($r = 0.28$, $p = 0.029$), primarily driven by a higher rate of out-context errors (i.e., applying outdated rules to new contexts). This suggests stress may impair rule updating and feature selection under dynamic conditions.

### 6.2 Mission 2: Instrumental Learning and Hierarchical Control

In Mission 2, participants were required to learn the higher-order rule through experience. Specifically, they had to link different cue types (cars) to the relevant code feature to predict the correct response (reward package drop-off). While learning curves were positive across participants (Fig. 4B), those who had shown greater flexibility in Mission 1 exhibited significantly higher learning rates in Mission 2, particularly as cue complexity increased across blocks ($r = 0.31$, $p = 0.015$). Consistent with Mission 1, higher self-reported stress predicted lower learning rates ($r = -0.30$, $p = 0.020$). Interestingly, self-reported depressed mood, which was unrelated to Mission 1 performance, was also associated with impaired learning ($r = -0.32$, $p = 0.010$), suggesting a unique link between affective symptoms and adaptive learning in uncertain environments.

### 6.3 Mission 3: Social Avoidance and Trust Adaptation

Mission 3 introduced a social component, where participants could delegate the code-solving task to partners with varying reliability and intent (kind, clumsy, jerk). Participants had to learn whether to rely on "kind," "clumsy," or "jerk" partners. Specifically, they had to learn the type of the different partners: kind partners were helpful, clumsy partners were error-prone but well-intentioned, and jerk partners were strategically untrustworthy.

Fig. 4C (top) shows that participants successfully learned to differentiate among the different partner types over time. Notably, participants with higher learning rates in Mission 2 were more likely to trust clumsy (but well-meaning) partners in Mission 3 ($r = 0.36$, $p = 0.004$), suggesting greater nuance in evaluating others' intent versus outcome. However, learning rates in mission 2 were not per se related to those in mission 3, suggesting that these missions tap into distinct learning mechanisms.

Self-efficacy also modulated partner-related decision-making. Specifically, participants with higher self-efficacy scores were initially more likely to double-check feedback from jerk partners ($r = 0.28$, $p = 0.031$), but later became less likely to follow any partners signals when their control was externally removed. Hence, they preferred to randomly guess independently rather than relying on enforced partnership ($r = -0.32$, $p = 0.011$). This behavioral shift suggests an interaction between perceived agency and contextual controllability. In later blocks, when partner interference was harder to avoid, many participants adapted by preemptively speeding up to maintain control, anticipating loss of agency and strategically avoiding reliance on unreliable others (Fig. 4C, bottom: comparing first full and second partial control blocks). These findings illustrate flexible strategy use in a socially complex and noisy environment.

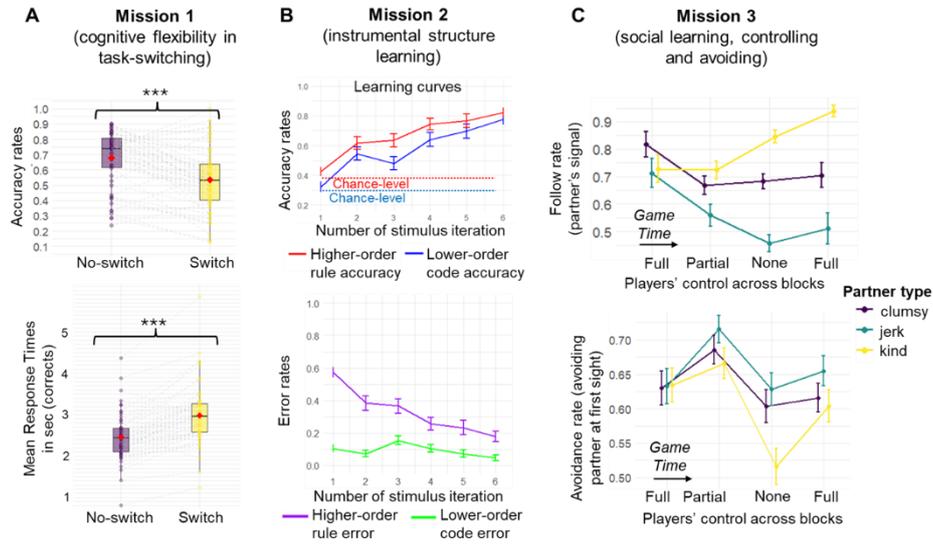

**Fig. 4.** Behavioral game patterns across the first three missions. **A.** Switch costs in Mission 1 show that we replicate classical findings from laboratory task-switching paradigms. Horizontal thick bars refer to medians; red diamonds refer to means; vertical bars refer to within-subject SEMs; asterisks refer to statistical significance (p < 0.001). *Top:* Lower-order code accuracy *Bottom:* Mean RTs of correct choices. **B.** Learning curves over time show that we replicate classical findings from laboratory instrumental learning tasks. *Top:* accuracy by number of times a given cue-code association was shown that had to be learned. Higher-order rule accuracy means that subjects correctly identified the relevant code feature (based on car cue) while lower-order rule accuracy means that subjects correctly classified the code stimulus based on the relevant feature. *Bottom:* Error patterns over time shows that people learned the relevant code feature over time but made lower-order code accuracy mistakes (because of perceptual difficulty). **C.** Participants learned to make their trust contingent on partner type (*top*) and adjust their strategy selectively towards them based on controllability shifts across blocks (*bottom*). *Top*: over time, participants learned to trust kind partners more than jerk partners. *Bottom*: in the second (partial control), participants lost control over code cracking if they let partners squeeze in front of them. This increased avoiding them at first sight, suggesting that participants adjusted their reliance on others accordingly.

### 6.4 Engagement, Dropout, and Ongoing Work

Despite the layered cognitive demands, dropout rates were low (< 2%), and participant feedback indicated high engagement and perceived value. This supports the feasibility of GF as a naturalistic yet computationally tractable paradigm for studying adaptability in a fun and engaging environment without losing scientific rigor. Current work is extending these analyses with hierarchical Bayesian models to quantify constructs such as effort sensitivity, controllability preferences, and learning asymmetries. These model-derived parameters will be related to individual differences in real-world functioning and clinical symptom profiles. Future studies are validating these mechanisms in clinical populations (e.g., remitted depression) and deploying the platform for longitudinal assessment to capture dynamic changes in adaptability over time.

## 7 AI Integration and Adaptive Feedback Loop

As AI systems increasingly aim to emulate human adaptability, there is a growing need for environments that challenge both humans and machines in structured, dynamic, and ecologically valid ways [1, 15]. Grounded in the Supertask paradigm, GF offers a high-dimensional, evolving testbed that can be shared by human players and artificial agents alike so that interactions lead to co-evolution. Specifically, through AI integration, GF is evolving into a closed-loop cognitive environment where human behavior, intelligent agents, and adaptive task structures co-develop [15]. This framework creates a novel platform for studying human-AI interaction, developing personalized learning tools, and testing the boundaries of both biological and artificial adaptability under shared constraints. As next steps, we are extending GF's capabilities by embedding artificial agents across three complementary dimensions: 1. benchmarking adaptability; 2. enabling personalized learning through latent modeling; and 3. supporting reflective gameplay via embedded conversational agents.

### 7.1 Benchmarking Human vs Artificial Adaptability

Artificial agents, ranging from Q-learners to meta-reinforcement learning models, can be exposed to the same sequence of missions as human players [65]. This enables direct comparisons of exploration strategies, context inference, effort-reward trade-offs, and generalization across tasks. These benchmarks provide a new lens on human adaptability, revealing both where it excels and where current AI systems fall short - especially in handling shifting goals, ambiguous feedback, or hidden task structure [4, 28, 46].

### 7.2 Adaptive Personalization Through Computational Modeling

The GF platform is built on a customized open-source jsPsych [47, 63] backend that is optimized for structured data collection and real-time adaptation. The front end is cross-platform compatible (desktop and tablet) and has been usability-tested in both research and clinical settings. Planned extensions include the integration of adaptive AI agents trained on player data using reinforcement learning and meta-learning. These agents can act as generative models of players' behavior by inferring latent cognitive states such as disengagement, uncertainty, or fatigue. This inference can then be used to dynamically adapting gameplay to support individualized learning trajectories. This layered design enables personalized feedback, including adjustments in tone, pacing, or challenge level, while maintaining the interpretability required for scientific and clinical insight.

### 7.3   Self-Reflection and Learning-by-Teaching

Future versions of GF will incorporate interactive game characters designed to reflect players' behavioral patterns and internal states. Framed as in-game drivers, these agents evolve through player choices, creating a psychologically immersive and self-relevant gameplay experience. Players will be prompted to teach coping strategies or decision heuristics to their avatars, transforming gameplay into an opportunity for structured self-instruction. This "learning-by-teaching" paradigm has demonstrated strong potential in educational robotics, where instructing humanoid agents led to improved learning outcomes and metacognitive awareness [40].

## 8   Discussion and Future Directions

Despite growing enthusiasm for serious games across science, healthcare, and education, many existing platforms fall short in at least four areas. First, they are often not designed with the end user in mind; particularly in clinical and institutional settings, where engagement is low and dropout rates are high [7, 41]. Second, many lack feedback mechanisms that feel meaningful to users, limiting both perceived value and opportunities for sustained reflection [32, 53]. Third, most fail to account for fluctuations in internal state or context, implicitly treating players as static learners rather than dynamic agents navigating changing demands [3, 15, 56]. Lastly, these platforms are rarely optimized for neurocomputational models that explain emerging behavioral patterns across analytical levels – models that are essential for identifying mechanisms of behavior change, personalizing interventions, and reconfiguring neurobiological markers to support recovery in conditions like depression, anxiety, and ADHD [30].

GF addresses these limitations by introducing a new Supertask paradigm, combining serious gaming, neurocomputational modeling, behavioral economic theory, and AI. The platform unites three traditionally fragmented goals: assessing neurocomputational mechanisms of adaptability, informing personalized clinical models, and helping individuals and agents to co-evolve and learn together to adapt across shifting contexts. Its layered architecture supports computational modeling across multiple timescales while preserving intuitive, mission-based gameplay that remains engaging and ecologically valid.

This integrative framework has broad implications. In computational psychiatry, GF provides a scalable platform for digital phenotyping, capturing fluctuations in cognitive control, strategy flexibility, and perceived controllability. These dimensions are often missed in static neuropsychological task batteries. Clinicians can leverage these insights to understand mechanisms of action behind digital interventions, determine when to initiate or discontinue specific treatments, and tailor therapeutic decisions based on the most malleable mechanisms of adaptation. In neuroscience and cognitive science, it functions as a dynamic testbed for modeling latent structure inference and adaptive behavior across nested temporal levels. In artificial intelligence, it offers a unified environment where agents can be directly benchmarked against humans, opening new opportunities for advancing generalization, meta-learning, and human-aligned decision-making. For users themselves, GF enables a meaningful and reflective gameplay experience, incorporating mechanics such as adaptive feedback and learning-by-teaching to promote insight, agency, and self-guided growth.

The present work establishes the Supertask paradigm as a conceptual and technical foundation for studying adaptability in cognitively and socially rich environments. GF serves as its first prototype, illustrating how serious games can be computationally structured to elicit and model adaptive mechanisms across multiple timescales. While we present initial construct-validating results, our aim here is not to provide a full empirical or clinical account. Instead, this work anchors a growing research agenda that will include formal modeling, longitudinal data collection, and clinical application. Future studies will expand upon this foundation to develop individualized neurocognitive and learning profiles, test dynamic intervention strategies, and benchmark adaptive capacity in humans and artificial agents that together co-evolve in a unified environment.

## 8.1 The Gearshift Fellowship Vision: Creating a New Integrative Ecosystem

*What if adaptability could be measured, modeled, and trained – across minds and machines?* Our vision is to create an ecosystem that uncovers latent adaptive mechanisms within a controlled yet ecologically valid testbed; a clinical tool for tracking and guiding mental health trajectories; a personalized learning engine for reflective, adaptive training; and a benchmarking platform for comparing human and artificial agents under shared task constraints. This ecosystem not only integrates advanced computational models with serious gaming but also incorporates diverse stakeholders (clinicians, researchers, and individuals) fostering a collaborative environment for innovation.

GF is not a tool – it is a platform for understanding adaptation as an ongoing process, not merely an outcome. This marks a step toward a future where technology helps us not only measure behavior but understand, shape, and align it with our evolving goals, for both humans and machines. By fostering the co-evolution of adaptive artificial agents and humans, GF aims to enhance consciousness and cognitive functions, building toward systems that continuously evolve to support both domains. This co-evolution not only enhances human cognition but also advances AI systems that learn and adapt from human interaction, leading to more intelligent, responsive agents capable of shaping their own behaviors and improving their understanding of human cognition.

By providing real-time, adaptive feedback, GF has the potential to revolutionize personalized interventions, enabling clinicians to tailor treatments to the dynamic needs of their patients and accelerating the adoption of computational psychiatry tools in clinical practice. Integrating modeling, design, and meaning-making, GF lays the foundation for a new class of neurocognitive environments. Namely, systems that support players, clinicians, researchers, and machines in better understanding how to adapt to a complex, uncertain world. While the path ahead is challenging and complex, GF is positioned as a scalable platform that evolves with the advances in both human cognitive science and AI development. Each phase of the project builds on the previous one, leading us closer to fully integrated systems that can shape behavior across human and artificial agents.

**Acknowledgments.** NGJ thanks Michael J. Frank for helpful discussions and encouragement. Deep gratitude goes to Donja Darai, Andrew Westbrook, Meghan Gallo, and all members of the LNCC lab at Brown University for their generous support and thoughtful feedback. We are also immensely grateful to our Beta User Community for their insights during testing. Additional thanks go to Seik Oh, Fiona Griffith, Nada Saaida, Macfadyen Nichols, and Chaeree Lee for valuable feedback on experimental design and user experience. Special thanks to Rachel Zhou for her outstanding design work, and to Paul Xu and the entire CCV engineering team at Brown University for their technical support in developing the Supertask prototype. All of these contributions were instrumental in shaping the direction and development of this project. This work was supported by grant P500PS_214223 from the Swiss National Science Foundation [to NGJ] and an ARC program award from the Carney Institute for Brain Science at Brown University [to NGJ].

**Author Contribution Statement.** NGJ conceived and led the project, developed the scientific framework, designed the task architecture, supervised the implementation, conducted all analyses, and wrote the manuscript. NGJ also secured the funding that enabled the development of the platform and supported project personnel. RKC led the technical development of the platform backend, contributed to system design and integration, and collaborated on aligning the platform with long-term translational objectives. JL and RG developed and implemented core software modules. RZ created the visual design elements and narrative components of the game environment.

**Disclosure of Interests.** The authors have no competing interests to declare that are relevant to the content of this article.

# References

1. Allen, K. et al.: Using games to understand the mind. Nat Hum Behav. 8, 6, 1035–1043 (2024). https://doi.org/10.1038/s41562-024-01878-9.
2. American Psychiatric Association: Diagnostic and Statistical Manual of Mental Disorders, 5th Edition: DSM-5. American Psychiatric Publishing, Washington, D.C (2013).
3. Birk, M.V., Mandryk, R.L.: Combating Attrition in Digital Self-Improvement Programs using Avatar Customization. In: Proceedings of the 2018 CHI Conference on Human Factors in Computing Systems. pp. 1–15 Association for Computing Machinery, New York, NY, USA (2018). https://doi.org/10.1145/3173574.3174234.
4. Botvinick, M. et al.: Reinforcement Learning, Fast and Slow. Trends in Cognitive Sciences. 23, 5, 408–422 (2019). https://doi.org/10.1016/j.tics.2019.02.006.
5. Botvinick, M.M. et al.: Effort discounting in human nucleus accumbens. Cognitive, Affective, & Behavioral Neuroscience. 9, 1, 16–27 (2009). https://doi.org/10.3758/CABN.9.1.16.
6. Botvinick, M.M.: Hierarchical reinforcement learning and decision making. Current Opinion in Neurobiology. 22, 6, 956–962 (2012). https://doi.org/10.1016/j.conb.2012.05.008.
7. Boyle, E.A. et al.: An update to the systematic literature review of empirical evidence of the impacts and outcomes of computer games and serious games.

Computers & Education. 94, 178–192 (2016). https://doi.org/10.1016/j.compedu.2015.11.003.
8. Brockman, G. et al.: OpenAI Gym, http://arxiv.org/abs/1606.01540, (2016). https://doi.org/10.48550/arXiv.1606.01540.
9. Cavanagh, J.F., Castellanos, J.: Identification of canonical neural events during continuous gameplay of an 8-bit style video game. NeuroImage. 133, 1–13 (2016). https://doi.org/10.1016/j.neuroimage.2016.02.075.
10. Chen, G. et al.: Validation of a New General Self-Efficacy Scale. Organizational Research Methods. 4, 1, 62–83 (2001). https://doi.org/10.1177/109442810141004.
11. Chevalier-Boisvert, M. et al.: BabyAI: A Platform to Study the Sample Efficiency of Grounded Language Learning, http://arxiv.org/abs/1810.08272, (2019). https://doi.org/10.48550/arXiv.1810.08272.
12. Cobbe, K. et al.: Quantifying Generalization in Reinforcement Learning. In: Proceedings of the 36th International Conference on Machine Learning. pp. 1282–1289 PMLR (2019).
13. Cohen, J.D. et al.: Should I stay or should I go? How the human brain manages the trade-off between exploitation and exploration. Philosophical Transactions of the Royal Society B: Biological Sciences. 362, 1481, 933–942 (2007). https://doi.org/10.1098/rstb.2007.2098.
14. Collins, A.G.E., Frank, M.J.: Cognitive control over learning: Creating, clustering, and generalizing task-set structure. Psychological Review. 120, 1, 190–229 (2013). https://doi.org/10.1037/a0030852.
15. Collins, K.M. et al.: Building machines that learn and think with people. Nat Hum Behav. 8, 10, 1851–1863 (2024). https://doi.org/10.1038/s41562-024-01991-9.
16. Conners, C. et al.: Conners' Adult ADHD Rating Scales (CAARS) technical manual. Multi-Health Systems, North Tonawanda, NY (1999).
17. Cools, R. et al.: Defining the Neural Mechanisms of Probabilistic Reversal Learning Using Event-Related Functional Magnetic Resonance Imaging. J. Neurosci. 22, 11, 4563–4567 (2002). https://doi.org/10.1523/JNEUROSCI.22-11-04563.2002.
18. Cooper, S. et al.: Predicting protein structures with a multiplayer online game. Nature. 466, 7307, 756–760 (2010). https://doi.org/10.1038/nature09304.
19. Coutrot, A. et al.: Global Determinants of Navigation Ability. Current Biology. 28, 17, 2861-2866.e4 (2018). https://doi.org/10.1016/j.cub.2018.06.009.
20. Daw, N.D. et al.: Model-Based Influences on Humans' Choices and Striatal Prediction Errors. Neuron. 69, 6, 1204–1215 (2011). https://doi.org/10.1016/j.neuron.2011.02.027.
21. Daw, N.D. et al.: Uncertainty-based competition between prefrontal and dorsolateral striatal systems for behavioral control. Nat Neurosci. 8, 12, 1704–1711 (2005). https://doi.org/10.1038/nn1560.
22. Diamond, A.: Executive Functions. Annual Review of Psychology. 64, Volume 64, 2013, 135–168 (2013). https://doi.org/10.1146/annurev-psych-113011-143750.

23. Din, S.U. et al.: Serious Games: An Updated Systematic Literature Review, http://arxiv.org/abs/2306.03098, (2023). https://doi.org/10.48550/arXiv.2306.03098.
24. Finn, C. et al.: Model-Agnostic Meta-Learning for Fast Adaptation of Deep Networks. In: Proceedings of the 34th International Conference on Machine Learning. pp. 1126–1135 PMLR (2017).
25. Forstmann, B.U. et al.: Sequential Sampling Models in Cognitive Neuroscience: Advantages, Applications, and Extensions. Annual Review of Psychology. 67, 1, 641–666 (2016). https://doi.org/10.1146/annurev-psych-122414-033645.
26. Frank, M.J.: Adaptive Cost-Benefit Control Fueled by Striatal Dopamine. (2025). https://doi.org/10.1146/annurev-neuro-112723-025228.
27. Gershman, S.J., Daw, N.D.: Reinforcement Learning and Episodic Memory in Humans and Animals: An Integrative Framework. Annual Review of Psychology. 68, Volume 68, 2017, 101–128 (2017). https://doi.org/10.1146/annurev-psych-122414-033625.
28. Gershman, S.J., Niv, Y.: Learning latent structure: carving nature at its joints. Current Opinion in Neurobiology. 20, 2, 251–256 (2010). https://doi.org/10.1016/j.conb.2010.02.008.
29. Ging-Jehli, N. et al.: Dissecting Neurocomputational Mechanisms of Impaired Instrumental Learning Across Psychopathologies Using Integrative Model-Based EEG Phenotyping. Biological Psychiatry. 97, 9, S7 (2025).
30. Ging-Jehli, N.R. et al.: Improving neurocognitive testing using computational psychiatry—A systematic review for ADHD. Psychological Bulletin. 147, 2, 169–231 (2021). https://doi.org/10.1037/bul0000319.
31. Ging-Jehli, N.R., Ratcliff, R.: Effects of aging in a task-switch paradigm with the diffusion decision model. Psychology and Aging. 35, 6, 850–865 (2020). https://doi.org/10.1037/pag0000562.
32. Granic, I. et al.: The benefits of playing video games. American Psychologist. 69, 1, 66–78 (2014). https://doi.org/10.1037/a0034857.
33. Griffiths, T.L. et al.: Bayesian models of cognition. In: The Cambridge handbook of computational psychology. pp. 59–100 Cambridge University Press, New York, NY, US (2008). https://doi.org/10.1017/CBO9780511816772.006.
34. Griffiths, T.L. et al.: Doing more with less: meta-reasoning and meta-learning in humans and machines. Current Opinion in Behavioral Sciences. 29, 24–30 (2019). https://doi.org/10.1016/j.cobeha.2019.01.005.
35. Hampshire, A. et al.: Fractionating Human Intelligence. Neuron. 76, 6, 1225–1237 (2012). https://doi.org/10.1016/j.neuron.2012.06.022.
36. Hardy, J.L. et al.: Enhancing Cognitive Abilities with Comprehensive Training: A Large, Online, Randomized, Active-Controlled Trial. PLOS ONE. 10, 9, e0134467 (2015). https://doi.org/10.1371/journal.pone.0134467.
37. Henry, J.D., Crawford, J.R.: The short-form version of the Depression Anxiety Stress Scales (DASS-21): Construct validity and normative data in a large non-clinical sample. British Journal of Clinical Psychology. 44, 2, 227–239 (2005). https://doi.org/10.1348/014466505X29657.

38. Huys, Q.J.M. et al.: Advances in the computational understanding of mental illness. Neuropsychopharmacol. 46, 1, 3–19 (2021). https://doi.org/10.1038/s41386-020-0746-4.
39. Huys, Q.J.M. et al.: Computational psychiatry as a bridge from neuroscience to clinical applications. Nat Neurosci. 19, 3, 404–413 (2016). https://doi.org/10.1038/nn.4238.
40. Jamet, F. et al.: Learning by Teaching with Humanoid Robot: A New Powerful Experimental Tool to Improve Children's Learning Ability. Journal of Robotics. 2018, e4578762 (2018). https://doi.org/10.1155/2018/4578762.
41. Kato, P.M.: Video Games in Health Care: Closing the Gap. Review of General Psychology. 14, 2, 113–121 (2010). https://doi.org/10.1037/a0019441.
42. Kiesel, A. et al.: Control and interference in task switching—A review. Psychological Bulletin. 136, 5, 849–874 (2010). https://doi.org/10.1037/a0019842.
43. Krath, J., Von Korflesch, H.F.: Designing gamification and persuasive systems: a systematic literature review. GamiFIN. 100–109 (2021).
44. Kumar, S. et al.: Meta-Learning of Structured Task Distributions in Humans and Machines, http://arxiv.org/abs/2010.02317, (2021). https://doi.org/10.48550/arXiv.2010.02317.
45. Laird, J.E. et al.: A Standard Model of the Mind: Toward a Common Computational Framework across Artificial Intelligence, Cognitive Science, Neuroscience, and Robotics. AI Magazine. 38, 4, 13–26 (2017). https://doi.org/10.1609/aimag.v38i4.2744.
46. Lake, B.M. et al.: Building machines that learn and think like people. Behavioral and Brain Sciences. 40, e253 (2017). https://doi.org/10.1017/S0140525X16001837.
47. Leeuw, J.R. de et al.: jsPsych: Enabling an Open-Source Collaborative Ecosystem of Behavioral Experiments. Journal of Open Source Software. 8, 85, 5351 (2023). https://doi.org/10.21105/joss.05351.
48. Lieder, F., Griffiths, T.L.: Resource-rational analysis: Understanding human cognition as the optimal use of limited computational resources. Behavioral and Brain Sciences. 43, e1 (2020). https://doi.org/10.1017/S0140525X1900061X.
49. Lin, J., Chang, W.-R.: Effectiveness of Serious Games as Digital Therapeutics for Enhancing the Abilities of Children With Attention-Deficit/Hyperactivity Disorder (ADHD): Systematic Literature Review. JMIR Serious Games. 13, 1, e60937 (2025). https://doi.org/10.2196/60937.
50. Loijen, A. et al.: Biased approach-avoidance tendencies in psychopathology: A systematic review of their assessment and modification. Clinical Psychology Review. 77, 101825 (2020). https://doi.org/10.1016/j.cpr.2020.101825.
51. Marne, B. et al.: The Six Facets of Serious Game Design: A Methodology Enhanced by Our Design Pattern Library. In: Ravenscroft, A. et al. (eds.) 21st Century Learning for 21st Century Skills. pp. 208–221 Springer, Berlin, Heidelberg (2012). https://doi.org/10.1007/978-3-642-33263-0_17.
52. Miller, E.K., Cohen, J.D.: An Integrative Theory of Prefrontal Cortex Function. Annual Review of Neuroscience. 24, Volume 24, 2001, 167–202 (2001). https://doi.org/10.1146/annurev.neuro.24.1.167.

GF-Adaptability     21

53. Mitsea, E. et al.: A Systematic Review of Serious Games in the Era of Artificial Intelligence, Immersive Technologies, the Metaverse, and Neurotechnologies: Transformation Through Meta-Skills Training. Electronics. 14, 4, 649 (2025). https://doi.org/10.3390/electronics14040649.
54. Morris, L., Mansell, W.: A systematic review of the relationship between rigidity/flexibility and transdiagnostic cognitive and behavioral processes that maintain psychopathology. Journal of Experimental Psychopathology. 9, 3, 2043808718779431 (2018). https://doi.org/10.1177/2043808718779431.
55. Onnela, J.-P., Rauch, S.L.: Harnessing Smartphone-Based Digital Phenotyping to Enhance Behavioral and Mental Health. Neuropsychopharmacol. 41, 7, 1691–1696 (2016). https://doi.org/10.1038/npp.2016.7.
56. Parsons, T.D.: Virtual Reality for Enhanced Ecological Validity and Experimental Control in the Clinical, Affective and Social Neurosciences. Front. Hum. Neurosci. 9, (2015). https://doi.org/10.3389/fnhum.2015.00660.
57. Pistono, A.M.A.D.A. et al.: A Review of Adaptable Serious Games Applied to Professional Training. Journal of Digital Media & Interaction. 4 n.º 11, 60-85 Páginas (2021). https://doi.org/10.34624/JDMI.V4I11.26419.
58. Ratcliff, R.: A theory of memory retrieval. Psychological Review. 85, 2, 59–108 (1978). https://doi.org/10.1037/0033-295X.85.2.59.
59. Rhoads, S.A. et al.: Advancing computational psychiatry through a social lens. Nat. Mental Health. 2, 11, 1268–1270 (2024). https://doi.org/10.1038/s44220-024-00343-w.
60. Rosser, B.A.: Intolerance of Uncertainty as a Transdiagnostic Mechanism of Psychological Difficulties: A Systematic Review of Evidence Pertaining to Causality and Temporal Precedence. Cogn Ther Res. 43, 2, 438–463 (2019). https://doi.org/10.1007/s10608-018-9964-z.
61. Shea, N. et al.: Supra-personal cognitive control and metacognition. Trends in Cognitive Sciences. 18, 4, 186–193 (2014). https://doi.org/10.1016/j.tics.2014.01.006.
62. Shenhav, A. et al.: The Expected Value of Control: An Integrative Theory of Anterior Cingulate Cortex Function. Neuron. 79, 2, 217–240 (2013). https://doi.org/10.1016/j.neuron.2013.07.007.
63. Strittmatter, Y. et al.: A jsPsych touchscreen extension for behavioral research on touch-enabled interfaces. Behav Res. 56, 7, 7814–7830 (2024). https://doi.org/10.3758/s13428-024-02454-9.
64. Talebi, M. et al.: The Cambridge Neuropsychological Test Automated Battery (CANTAB) Versus the Minimal Assessment of Cognitive Function in Multiple Sclerosis (MACFIMS) for the Assessment of Cognitive Function in Patients with Multiple Sclerosis. Multiple Sclerosis and Related Disorders. 43, 102172 (2020). https://doi.org/10.1016/j.msard.2020.102172.
65. Wang, J.X. et al.: Prefrontal cortex as a meta-reinforcement learning system. Nat Neurosci. 21, 6, 860–868 (2018). https://doi.org/10.1038/s41593-018-0147-8.
66. Wiecki, T.V. et al.: Model-based cognitive neuroscience approaches to computational psychiatry: Clustering and classification. Clinical Psychological Science. 3, 3, 378–399 (2015). https://doi.org/10.1177/2167702614565359.

67. Yu, T. et al.: Meta-World: A Benchmark and Evaluation for Multi-Task and Meta Reinforcement Learning, http://arxiv.org/abs/1910.10897, (2021). https://doi.org/10.48550/arXiv.1910.10897.
68. BrainHQ from Posit Science, https://www.brainhq.com/, last accessed 2025/05/24.
69. Effectiveness of Remote Cognitive Assessment: Examining Results from the Healthy Brain Project, https://www.cogstate.com/blog/remote-cognitive-assessment-examining-results-from-the-healthy-brain-project/, last accessed 2025/05/24.